\documentclass[12pt]{article}
\usepackage{amsmath}
\usepackage[noend]{algpseudocode}
\usepackage{algorithm2e}
\usepackage{amssymb}
\usepackage{graphicx}
\usepackage{psfrag}
\usepackage{hyperref}
\usepackage{tikz}
\usepackage{tikz-cd}

\newcommand{\reals}{{\mbox{\bf R}}}
\newcommand{\integers}{{\mbox{\bf Z}}}

\newcommand{\ie}{{\it i.e.}}

\newcommand{\ones}{\mathbf 1}

\newcommand{\BIT}{\begin{itemize}}
\newcommand{\EIT}{\end{itemize}}
\newcommand{\BEQ}{\begin{equation}}
\newcommand{\EEQ}{\end{equation}}
\newcommand{\BEAS}{\begin{eqnarray*}}
\newcommand{\EEAS}{\end{eqnarray*}}

\newcounter{algorithmctr}[section]
\renewcommand{\thealgorithmctr}{\thesection.\arabic{algorithmctr}}
{\refstepcounter{algorithmctr}
\begin{list}{}{%
		\setlength{\rightmargin}{0\linewidth}%
		\setlength{\leftmargin}{.05\linewidth}}%
	\rmfamily\small
	\item[]{\setlength{\parskip}{0ex}\hrulefill\par%
		\nopagebreak{\bfseries\textsf{Algorithm \thealgorithmctr~}}}}%
{{\setlength{\parskip}{-1ex}\nopagebreak\par\hrulefill}
\end{list}}

{\refstepcounter{algorithmctr}
\begin{list}{}{%
		\setlength{\rightmargin}{0\linewidth}%
		\setlength{\leftmargin}{.05\linewidth}}%
	\rmfamily\small
	\item[]{\setlength{\parskip}{0ex}\hrulefill\par%
		\nopagebreak{\bfseries\textsf{Algorithm}}}}%
{{\setlength{\parskip}{-1ex}\nopagebreak\par\hrulefill}
\end{list}}

\begin{document}

\title{Efficient Shapley Performance Attribution\\ for Least-Squares
	Regression}
\author{Logan Bell \and Nikhil Devanathan \and Stephen Boyd}
\date{\today}
\maketitle

\begin{abstract}
We consider the performance of a least-squares regression model, as
judged by out-of-sample $R^2$. Shapley values give a fair attribution of the
performance of a model to its input features, taking into account
interdependencies between features. Evaluating the Shapley values exactly
requires solving a number of regression problems that is exponential in the
number of features, so a Monte Carlo-type approximation is typically used. We
focus on the special case of least-squares regression models, where several
tricks can be used to compute and evaluate regression models efficiently.
These tricks give a substantial speed up, allowing many more Monte Carlo
samples to be evaluated, achieving better accuracy. We refer to our method as
least-squares Shapley performance attribution (LS-SPA), and describe our
open-source implementation.
\end{abstract}

This version of the article has been accepted for publication, after peer review 
and is subject to Springer Nature’s AM terms of use, but is not the Version of 
Record and does not reflect post-acceptance improvements, or any corrections. The 
Version of Record is available online at: 
\url{http://dx.doi.org/10.1007/s11222-024-10459-9}.

\clearpage
\section{Introduction}\label{s-intro}

We consider classic least-squares regression, with $p$ features, judged by an
out-of-sample $R^2$ metric. A natural question is how much each of the $p$
features contributes to our $R^2$ metric; roughly speaking, how valuable is
each feature to our least-squares predictor? Except for a special case
described below in \S\ref{s-trivial}, this question seems difficult to answer,
since the value of a feature depends on the other features.

Our interest is in attributing the \emph{overall performance} of a
least-squares model to the features. A related task is attributing a
\emph{specific prediction} of a least-squares model to the features, which is a
popular method for so-called explainable AI called SHAP, an acronym for Shapley
additive explanations \cite{Lundberg2017,Molnar2022,Chen2023}. That is a very
different task, discussed in more detail below. In this paper, we consider only
performance attribution, and not explaining a specific prediction from a model.
We refer to this task as Shapley performance attribution to features.

This performance attribution problem was essentially solved in Lloyd Shapley's
1953 paper ``A Value for $n$-Person Games'' \cite{Shapley1952}. He proposed a
method to allocate the payoff in a cooperative game to the players, which came
to be known as Shapley values. Shapley values provide a fair
distribution of the total payoff in a game, taking into account the
contributions of each player to the coalition. Shapley values are provably
the only attribution for which fairness, monotonicity, and full attribution
(three key desiderata for attribution) all hold. We refer the reader to other
papers for more discussion and justification of Shapley values for attributing
regression model performance to its features
\cite{Huettner2012,Zhang2023,Fryer2021,Owen2017}.

We focus on efficiently computing (an approximation of) the Shapley values for
least-squares regression problems, \emph{i.e.}, to attribute the overall $R^2$
to the $p$ features. We seek a number $S_j$ associated with feature $j$, where
we interpret $S_j$ as the portion of the achieved $R^2$ metric that is
attributed to feature $j$. Full attribution means $\sum_{j=1}^p S_j = R^2$.

Shapley values rely on solving and evaluating around $2^p$ least-squares
problems. This is impractical for $p$ larger than around 10, so Monte Carlo
approximation is typically used to compute an approximation to the Shapley
values. We propose a simple but effective quasi-Monte Carlo method that in
practice gives better approximations of the Shapley values than Monte Carlo for
the same number of least-square regression problems.

We do not introduce any new mathematical or computational methods. Instead, we
collect well-known ideas and assemble them into an efficient method for
computing the Shapley values for a least-squares regression problem, exploiting
special properties of least-squares problems.

\subsection{Prior work}

\paragraph{Cooperative game theory} Shapley values originated in
cooperative game theory as a means of fairly splitting a coalition's reward
between the individual players \cite{Shapley1952}. The notion of a fair split
is defined by four axioms, which Shapley proved resulted in a unique method for
attribution. Since Shapley's seminal paper, numerous extensions, variations,
and generalizations have been developed; see, for instance, \cite{Monderer2002,
Dubey1981, Owen1977, Algaba2019, Chalkiadakis2012, Koczy2007}.

Computing the Shapley values in general has a cost that increases exponentially
in the number of players.  Nonetheless, many games have structure that enables
efficient exact computation of the Shapley values. Examples include weighted
hypergraph games with fixed coalition sizes \cite{Deng1994}, determining
airport landing costs \cite{Littlechild1973}, weighted voting games restricted
by trees \cite{Fernandez2002}, cost allocation problems framed as extended tree
games \cite{Granot2002}, sequencing games \cite{Curiel1989}, games represented
as marginal contribution networks \cite{Ieong2005}, and determining certain
notions of graph centrality \cite{Michalak2013E}. On the other hand, computing
Shapley values in weighted majority games is \#P-complete \cite{Deng1994},
as are elementary games, \emph{i.e.}, games whose value function is an
indicator on a coalition \cite{Faigle1992}.

\paragraph{Approximating Shapley values} Due to the computational
complexity of computing exact Shapley values in general, various methods have
been proposed for efficiently approximating Shapley values. Shapley initially
described a Monte Carlo method for approximating Shapley values by sampling
coalitions in 1960 \cite{Mann1960}. Subsequent works have considered sampling
permutations using simple Monte Carlo methods
\cite{Zlotkin1994,Castro2009,Moehle2022}, stratified and quasi-Monte Carlo
methods \cite{vanCampen2017,Castro2017,Maleki2014,Mitchell2022}, and ergodic
sampling methods \cite{Illes2022}.

Beyond Monte Carlo approaches, other works have explored numerical integration
schemes for approximating the Shapley values. The paper \cite{Owen1972}
describes a multilinear extension of the characteristic function of an
$n$-person game that allows for the computation of the Shapley values as a
contour integral. This method has been further explored in \cite{Leech2003} and
\cite{Fatima2008}.

\paragraph{Applications of Shapley values}
Although they arose in the context of game theory, Shapley values have been
applied across a variety of fields. In finance, Shapley values have been
applied to attribute the performance of a portfolio to constituent assets
\cite{Moehle2022} and to allocate insurance risk \cite{Powers2007}. Elsewhere,
Shapley values have been used to identify key individuals in social networks
\cite{Michalak2013,vanCampen2017}, to identify which components of a user
interface draw the most user engagement \cite{Zhao2018}, to distribute rewards
in multi-agent reinforcement learning \cite{Wang2020}, and to attribute the
performance of a machine-learning model to the individual training data points
\cite{Ghorbani2019}. We refer to \cite{Moretti2008} and \cite{Algaba2019} for a
deeper review of applications of Shapley values.

\paragraph{Explainable ML}
Shapley attribution has recently found extensive use in machine learning in the
context of model interpretability, in Shapley additive explanation (SHAP)
\cite{Lundberg2017}. SHAP uses approximate Shapley values to attribute a single
prediction of a machine-learning model across the input features.  Although
SHAP and Shapley performance attribution both involve prediction models and
both use Shapley values, they otherwise have little relation.  We refer to
\cite{Molnar2022} and \cite{Chen2023} for a more thorough review of SHAP.

\paragraph{Shapley values for statistics}
In statistical learning, researchers often seek to assign a relative importance
score to the features of a model. One approach is Shapley attribution. This
method has been independently rediscovered numerous times and called numerous
names \cite{Lindeman1980,Lipovetsky2001,
Kruskal1987,Mishra2016,Grmping2006,Grmping2015}. All of these works utilize
Shapley attribution to decompose the $R^2$ of a regression model, though often
without reference to Shapley. The paper \cite{Budescu1993} decomposes the $R^2$
using a method similar to Shapley attribution but with different weights, and
\cite{Chevan1991} decomposes any goodness-of-fit metric of a regression model
using a method shown in \cite{Stufken1992} to be equivalent to Shapley
attribution.

While not directly related to the computation of Shapley values, the
application of Shapley values to feature importance is a primary motivation
behind their calculation in many contexts
\cite{Moehle2022,Michalak2013,vanCampen2017}. In statistics, the use of Shapley
values for determining feature importance has been significantly explored
\cite{Kumar2020,Harris2022,Williamson2020,Fryer2021,Owen2017}, and papers
\cite{Huettner2012,Zhang2023,Fryer2021,Owen2017} further argue why Shapley
attribution is a particularly appropriate method for evaluating feature
importance.

\subsection{This paper}
We introduce an efficient method for approximating Shapley
attribution of performance in least-squares regression problems, called
least-squares Shapley performance attribution (LS-SPA).  LS-SPA uses several
computational tricks that exploit special properties of least-squares problems.
The first is a reduction of the original train and test data to a compressed
form in which the train and test data matrices are square. The second is to
solve a set of $p$ least-squares problems, obtained as we add features one by
one, with one QR factorization, in a time comparable to solving one
least-squares problem. Finally, we propose using a quasi-Monte Carlo method, a
variation of Monte Carlo sampling, to efficiently approximate the Shapley
values. (This trick does not depend on any special properties of least-squares
problems.)

\paragraph{Outline}
In \S\ref{s-ls-shapley} we present a mathematical overview of least-squares and
Shapley values, setting our notation. We describe our method for efficiently
estimating Shapley values for least-squares problems in \S\ref{s-compute}. In
\S\ref{s-ext}, we describe some extensions and variations on our algorithm, and
we conclude with numerical experiments in \S\ref{s-num}.

\section{Least-squares Shapley performance values}\label{s-ls-shapley}

In this section, we review the least-squares regression problem, set our
notation, and define the Shapley values for the features.

\subsection{Least-squares}\label{s-least-squares}
We consider the least-squares regression problem
\begin{equation}\label{e-lsq}
\begin{array}{ll}
\mbox{minimize} & \| X \theta - y\|_2^2,
\end{array}
\end{equation}
with variable $\theta\in \mathbf{R}^p$, the model parameter. Here $X \in
\reals^{N \times p}$ is a given data or feature matrix and $y \in \reals^N$ is
a given vector of responses or labels. The rows of $X$, denoted $x_i^T$ with
$x_i \in \reals^p$, correspond to $N$ samples or observations, and each column
of $X$ corresponds to a feature. We will assume that $X$ has rank $p$, which
implies $N \geq p$, \ie, $X$ is square or tall. We denote the solution of the
least-squares problem \eqref{e-lsq} as
\BEAS
\theta^\star = X^\dagger y = (X^TX)^{-1}X^Ty.
\EEAS
The data $X$ and $y$ are the training data since they are used to find the
model parameter $\theta^\star$.

The least-squares problem \eqref{e-lsq} yields a linear model $\hat
y=x^T\theta$ with $\theta=\theta^\star$.
We can include a constant offset or intercept in the model,
$\hat y = x^T\theta + \beta$, several ways.
One method is to include a feature that has the constant value one,
so the formulation above \eqref{e-lsq} is unchanged.
In this case, however, our attribution gives an attribution to the offset,
which might not be wanted.
Another method is to
solve \eqref{e-lsq} with centered data, \ie, data with the average of
each feature, and the labels, subtracted, so they all have zero mean.
To see this, note that the optimization problem
\[\begin{array}{ll}
\mbox{minimize} & \| X\theta + \beta\ones - y\|_2^2
\end{array}\]
with variables $\theta\in\reals^p$ and $\beta\in\reals$ has optimal
variables
\[
\theta^\star = (X-\ones\bar x)^\dagger(y -\ones\bar y), \quad
\beta^\star =
\frac{1}{N}\ones^T(y - X\theta^\star).
\]
Here, $\bar x=\frac{1}{N}\ones^T X$ is
the sample mean of the feature vectors and $\bar y=\frac{1}{N}\ones^T y$ is the
sample mean of the labels. Hence, to fit a model with an intercept, we can
solve \eqref{e-lsq} with centered data to obtain $\theta^\star$, and
from this recover $\beta^\star$.  In this formulation, we do not
attribute performance to the offset constant.
In the sequel we do not include the intercept, noting that an offset
can be included by centering the data.

\paragraph{Out-of-sample $R^2$ metric.}
We evaluate the performance of a model parameter $\theta$ via out-of-sample
validation. We have a second (test) data set of $M$ observations
$X^\mathrm{tst} \in \reals^{M \times p}$ and $y^\mathrm{tst} \in \reals^M$ and
evaluate the model on these data to obtain $\hat y^\mathrm{tst}=X^\mathrm{tst}
\theta$. The prediction errors on the test set are given by $\hat
y^\mathrm{tst} - y^\mathrm{tst}$. To evaluate the least-squares model with
parameter $\theta$, we use the $R^2$ metric
\BEQ\label{e-R2}
R^2 = \frac{
\|y^\mathrm{tst} \|_2^2 - \|\hat y^\mathrm{tst} - y^\mathrm{tst} \|_2^2}
{\|y^\mathrm{tst} \|_2^2},
\EEQ
which is the fractional reduction in mean square test error compared to the
baseline prediction $\hat y = 0$. Larger values of $R^2$ are better.  It is at
most one and can be negative.

In this paper, we focus exclusively on the out-of-sample $R^2$ metric
to evaluate a least-squares model.
However, the algorithm we develop and present in \S\ref{s-compute} can be
used to attribute any in-sample or out-of-sample performance metric
across the features of a least-squares model.

\subsection{Feature subsets and chains}\label{s-ftchain}
\paragraph{Feature subsets.}
In later sections, we will be interested in the $R^2$ metric obtained with the
least-squares model using only a subset $\mathcal S \subseteq \{1,\ldots, p\}$
of the features, \ie, using a parameter vector $\theta$ that satisfies
$\theta_j=0$ for $j \not\in \mathcal S$. The associated least-squares problem
is \BEQ\label{e-ftsub}
\begin{array}{ll}
\mbox{minimize} & \| X \theta - y\|_2^2\\
\mbox{subject to} & \theta_j = 0, \quad j \not\in \mathcal S.
\end{array}
\EEQ We denote the associated parameter as $\theta^\star_{\mathcal S}$. From
this we can find the $R^2$ metric, denoted $R^2_{\mathcal S}$, using
\eqref{e-R2}. We use $R^2$ to denote the metric obtained using all features,
\ie, $R^2_{\{1, \ldots, p\}}$.

\paragraph{Feature chains.}
A \emph{feature chain} is an increasing sequence of $p$ subsets of features
obtained by adding one feature at a time,
\BEAS
\emptyset \subset \mathcal S_1 \subset \cdots \subset
\mathcal S_p =\{1, \ldots, p\},
\EEAS
where $|\mathcal S_k|=k$. We denote $\pi_k$ as the index of the feature added
to form $\mathcal S_k$. Evidently $\pi=(\pi_1, \ldots, \pi_p)$ is a permutation
of $(1, \ldots, p)$. With this notation we have
\BEAS
\mathcal S_k =
(\pi_1, \ldots, \pi_k), \quad k=1,\ldots, p.
\EEAS
Roughly speaking, $\pi$ gives the order in which we add features in the feature
chain. We will set $\mathcal S_0=\emptyset$.

\paragraph{Lifts associated with a feature chain.}\label{s-lifts}
Consider feature $j$.  It is the $l$th feature to be added in the feature chain
given by $\pi$, where $l = \pi^{-1}(j)$. We define the \emph{lift} associated
with feature $j$ in chain $\pi$ as
\BEAS
L(\pi)_j = R^2_{\mathcal S_l} - R^2_{\mathcal S_{l-1}}.
\EEAS
Roughly speaking, $L(\pi)_j$ is the increase in $R^2$ obtained when we add
feature $j$ to the ones before it in the ordering $\pi$, \ie, features $\pi_1,
\ldots, \pi_{l-1}$. The lift $L(\pi)_j$ can be negative, which means that
adding feature $j$ to the ones that come before it reduces the $R^2$ metric.

We refer to the vector $L(\pi)\in \reals^p$ as the lift vector associated with
the feature chain given by $\pi$. We observe that
\BEAS
\sum_{j=1}^p L(\pi)_j =
\sum_{j=1}^p \left(  R^2_{\mathcal S_l} - R^2_{\mathcal S_{l-1}}\right) =
R^2,
\EEAS
the $R^2$ metric obtained using all features. The vector $L(\pi)$ gives an
attribution of the values of each feature to the final $R^2$ obtained, assuming
the features are added in the order $\pi$.  In general, it depends on $\pi$.


\subsection{Shapley attributions}\label{s-shapval} The vector of Shapley
attributions for the features, denoted $S\in \reals^p$, is given by
\BEQ\label{e-shapley-def} S = \frac{1}{p!} \sum_{\pi \in \mathcal P} L(\pi),
\EEQ where $\mathcal P$ is the set of all $p!$ permutations of $\{1,\ldots,
p\}$. We interpret $S_j$ as the average lift, or increase in $R^2$, obtained
when adding feature $j$ over all feature chains. The average is over all
feature chains, \ie, orderings of the features. In \S\ref{s-toy}, we
present a simple example of a Shapley attribution for a least-squares model
with a small number of features.

For $p$ more than 10 or so, it is impractical to evaluate the lift vector for
all $p!$ permutations. Instead, we estimate it as \BEQ\label{e-mc-approx} \hat
S = \frac{1}{K} \sum_{\pi \in \Pi} L(\pi), \EEQ where $\Pi \subset \mathcal P$
is a subset of permutations with $|\Pi|=K \ll p!$. This is a Monte Carlo
approximation of \eqref{e-shapley-def} when $\Pi$ is a subset of permutations
chosen uniformly at random from $\mathcal S$ with replacement.  (We will
describe a better choice in \S\ref{s-qmc}.)

\subsection{Uncorrelated features}\label{s-trivial} We mention here one case in
which the Shapley performance attribution for least-squares regression is
easily found: When the empirical covariance of the features on both the train
and test sets are diagonal, \ie,
\BEAS
(1/N) X^TX = \Lambda, \qquad  (1/M) (X^\mathrm{tst})^T X^\mathrm{tst} =
\tilde \Lambda,
\EEAS
with $\Lambda$ and $\tilde \Lambda$  diagonal. In this case, we have
$\theta_j^\star = \Lambda_{jj}^{-1}(X^Ty)_j$, for any subset $\mathcal S$ that
contains $j$. The test error is also additive, \ie, the sum of contributions
from each feature. It follows that the lift vectors do not depend on $\pi$, so
$S=L(\pi)$ for any $\pi$.

When these assumptions almost hold, \ie, the features are not too correlated on
the train and test sets, the method we propose exhibits very fast convergence.

\subsection{Toy example}\label{s-toy} To illustrate the ideas above we present
a simple example.  We use a synthetic dataset with $p=3$ features, $N=50$
training examples, and $M=50$ test examples. We generate feature matrices $X$
and $X^\text{tst}$ by taking, respectively, $N$ and $M$ independent samples from
a multivariate normal distribution with mean zero and covariance
\BEAS
\Sigma = \begin{bmatrix}
	1.0 &    0.7 &  -0.4\\
	0.7 &    1.0 &  -0.5\\
	-0.4 &  -0.5 &    1.0
\end{bmatrix}.
\EEAS
Using true weights $\theta=(2.1, 1.4, 0.1)$, we take $y=X\theta + \omega$ and
$y^\text{tst}=X^\text{tst}\theta+\omega^\text{tst}$ where the entries of
$\omega\in \reals^N$ and $\omega^\text{tst}\in \reals^M$ are independently
sampled from a standard normal distribution.

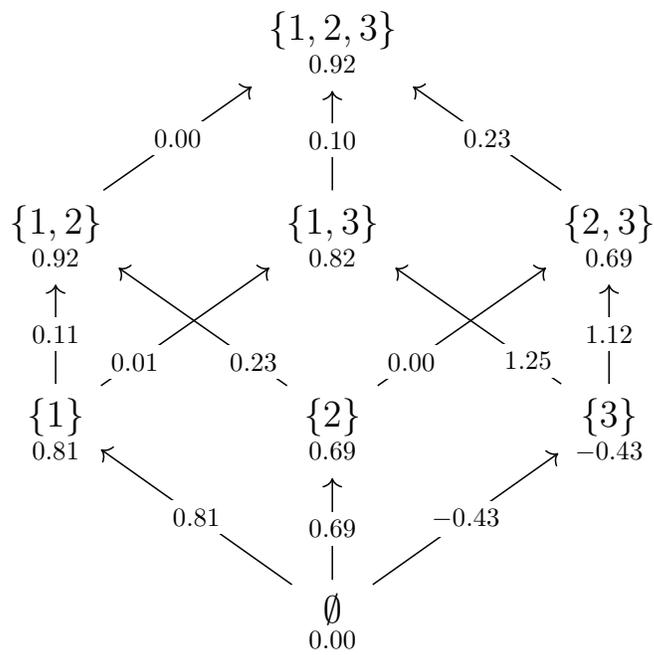
\begin{figure}
	\begin{large}
		\BEAS\begin{tikzcd}[sep=large] & {\underset{0.92}{{\{1,2,3\}}}} \\
			{\underset{0.92}{{\{1,2\}}}} & {\underset{0.82}{{\{1,3\}}}} &
			{\underset{0.69}{{\{2,3\}}}} \\
			{\underset{0.81}{{\{1\}}}} & {\underset{0.69}{{\{2\}}}} &
			{\underset{-0.43}{{\{3\}}}} \\
			& {\underset{0.00}{{\emptyset}}}
			\arrow["{-0.43}"{description}, from=4-2, to=3-3]
			\arrow["{0.69}"{description}, from=4-2, to=3-2]
			\arrow["{0.81}"{description}, from=4-2, to=3-1]
			\arrow["{0.11}"{description}, from=3-1, to=2-1]
			\arrow["{0.23}"{description, pos=0.2}, from=3-2, to=2-1]
			\arrow["{0.01}"{description, pos=0.2}, from=3-1, to=2-2]
			\arrow["{0.00}"{description, pos=0.2}, from=3-2, to=2-3]
			\arrow["{1.12}"{description}, from=3-3, to=2-3]
			\arrow["{0.10}"{description}, from=2-2, to=1-2]
			\arrow["{0.00}"{description}, from=2-1, to=1-2]
			\arrow["{0.23}"{description}, from=2-3, to=1-2]
			\arrow["{1.25}"{description, pos=0.2}, from=3-3, to=2-2]
		\end{tikzcd}\EEAS
	\end{large}
	\caption{Shapley attribution on the toy data represented as a lattice.}
	\label{f-lattice}
\end{figure}

Table~\ref{t-r2} shows the out-of-sample $R^2$ for each of the $8$ subsets of
features. Table~\ref{t-lift} shows the lift associated with each of $6$ feature
orderings. We display the same data as a lattice in figure~\ref{f-lattice}. In
this figure vertices are labeled with subsets of the features and subscripted
with the associated $R^2$. The edges, oriented to point to the subset to which
one feature was added, are labeled with the lift for adding that feature to the
subset. Every path from $\emptyset$ to $\{1,2,3\}$ corresponds to an ordering
of the features, with the lifts along the path giving the associated lift
vector.
\begin{table}
	\centering
	\begin{tabular}{c|r}
		$\mathcal{S}$ & $R^2$ \\
		\hline
		$\{1,2,3\}$ &  $0.92$\\
		$\{1,2\}$ &  $0.92$ \\
		$\{1,3\}$ &  $0.82$ \\
		$\{2,3\}$ &  $0.69$ \\
		$\{1\}$ &  $0.81$ \\
		$\{2\}$ &  $0.69$\\
		$\{3\}$ &  $-0.43$\\
		$\emptyset$ &  $0.00$
	\end{tabular}
	\caption{$R^2$ for each subset $\mathcal S$ of the features.}
	\label{t-r2}
\end{table}

\begin{table}
	\centering
	\begin{tabular}{c|rrr}
		$\pi$ & & $L(\pi)$ & \\
		\hline
		$(1,2,3)$ &  $(0.81,$&$ 0.11,$&$ 0.00)$\\
		$(1,3,2)$ &  $(0.81,$&$ 0.10,$&$ 0.01)$ \\
		$(2,1,3)$ &  $(0.23,$&$ 0.69,$&$ 0.00)$ \\
		$(2,3,1)$ &  $(0.23,$&$ 0.69,$&$ 0.00)$ \\
		$(3,1,2)$ &  $(1.25,$&$ 0.10,$&$ -0.43)$ \\
		$(3,2,1)$ &  $(0.23,$&$ 1.12,$&$ -0.43)$
	\end{tabular}
	\caption{Lift vector $L$ generated by each permutation $\pi$ of the features.}
	\label{t-lift}
\end{table}

The $R^2$ using all features is $0.92$, and the Shapley values are
\BEAS
S = (0.59,  0.47, -0.14).
\EEAS
Roughly speaking, most of our performance comes from feature~1, followed
closely by feature~2, with feature~3 negatively affecting performance.
(Since $R^2$ is evaluated out of sample, it can be negative.)
Indeed, we can see that the performance using only
features~1 and~2 is the same, to two
decimal places, as the performance using all three.

\section{Efficient computation}\label{s-compute} In this section we explain
LS-SPA, our method for efficiently computing $\hat S$, an approximation of $S$.
The method can be broken into two parts.  The first is a method to efficiently
compute $L(\pi)$, the lift associated with a specific feature ordering $\pi$.
The second is a method for choosing the set of permutations $\Pi$ that gives a
better approximation than basic Monte Carlo sampling.

\subsection{The na\"{i}ve method} The na\"{i}ve method for computing $\hat S$
is to solve a chain of $p$ least-squares problems $K$ times, and evaluate them
on a test set. Solving a least-squares problem with $k$ (nonzero) coefficients
has a cost $O(Nk^2)$ flops. (It can be done, for example, via the QR
factorization.) Evaluating its performance costs $O(Mk)$. Assuming $M$ is no
more than $Nk$ in order, this second term is negligible. Summing $O(Nk^2)$ from
$k=1$ to $p$ gives $O(Np^3)$. This is done for $K$ permutations so the
na\"{i}ve method requires \BEQ\label{e-complexity-naive} O(KNp^3) \EEQ flops.
This na\"{i}ve method can be parallelized: All of the least-squares problems
can be solved in parallel.

We will describe a method to carry out this computation far more efficiently.
The computation tricks we describe below are all individually well known; we
are merely assembling them into an efficient method.

\subsection{Initial reduction of training and test data sets}\label{s-reduce}

We can carry out an initial reduction of the original train and test data
matrices, so each has $p$ rows instead of $N$ and $M$ respectively. Let $X=QR$
denote the QR factorization of $X$, with $Q\in\reals^{N\times p}$ and $R\in
\reals^{p\times p}$. Simple algebra shows that \BEQ\label{e-reduction} \|
X\theta - y\|_2^2 =  \| R \theta - Q^Ty\|_2^2 + \|y - Q(Q^T y)\|_2^2. \EEQ The
righthand side consists of a least-squares objective with square data matrix
$R$ and righthand side $\tilde y=Q^Ty$, plus a constant. The cost to compute
$R$ and $\tilde y = Q^T y$ is $O(Np^2)$. We do this once and then solve the
least-squares problem \eqref{e-ftsub} using the objective $\|R \theta - \tilde
y\|_2^2$. The cost for this is $O(pk^2)$, where $k=|\mathcal S|$.

Computing the least-squares solutions for a chain now costs $O(p^4)$, whereas
in the na\"{i}ve method, the cost was $O(Np^3)$ per chain. The cost of
computing least-squares solutions for $K$ chains is then
\BEAS
O(Np^2 + Kp^4),
\EEAS
compared to $O(KNp^3)$ for the na\"{i}ve method. When $N$ or $K$ is large
(which is typical), the cost savings are substantial.

The same trick can be used to efficiently evaluate the $R^2$ metrics. We carry
out one QR factorization of the test matrix at a cost of $O(Mp^2)$, after which
we can evaluate the metric with $O(pk)$ flops, where $k=|\mathcal S|$.  To
evaluate the metrics for a chain is then $O(p^2)$ flops, compared to $O(Mp)$
for the na\"{i}ve method. To compute $\hat S$ for $K$ chains has cost
\BEAS
O(Mp^2 + Kp^2),
\EEAS
which is negligible compared to the cost of solving the least-squares problems.

Using this initial reduction trick, we obtain a complexity of $O(Np^2+Kp^4)$,
compared to $O(KNp^3)$ for the na\"{i}ve method. This simple trick has been
known since at least the 1960s \cite{Businger1965,Golub1965}. A similar
reduction trick uses a Cholesky factorization, rather than a QR factorization,
which is useful when the data set is too large to fit in memory; see
\S\ref{s-large-data} for more details.

\subsection{Efficiently computing lift vectors}\label{s-lift}

In this section, we show how the cost of computing lift vectors and evaluating
them for one chain can be reduced from $O(p^4)$ to $O(p^3)$, using a well-known
property of the QR factorization.

Given a feature chain $\mathcal S_0 \subset \mathcal S_1 \subset \cdots
\subset \mathcal S_p$ associated with a permutation $\pi$, solving
\eqref{e-ftsub} for $\mathcal S_i$ is equivalent to solving
\BEQ\label{e-ppsub}
\begin{array}{ll}
\mbox{minimize} & \| RP_\pi^T\tilde\theta - \tilde y\|_2^2\\
\mbox{subject to} & \tilde\theta_j = 0, \quad j > i,
\end{array}
\EEQ
with variable $\tilde\theta \in \reals^p$. Here $R$ and $\tilde y=Q^Ty$ are the
reduced data obtained from \S\ref{s-reduce} and $P_\pi$ is the permutation
matrix associated with $\pi$. The $k$th column of $RP_\pi^T$ is the $\pi_k$th
column of $R$. The optimal parameter $\theta^\star$ of \eqref{e-ftsub} is
related to the optimal parameter $\tilde\theta^\star$ of \eqref{e-ppsub} via
$\theta^\star = P_\pi^T\tilde\theta^\star$.

We can combine the problems \eqref{e-ppsub} for $i=1,\ldots,p$ into one problem
by collecting the parameter vectors $\tilde\theta$ into one $p \times p$ upper
triangular matrix $\tilde\Theta$. We then solve
\BEAS
\begin{array}{ll}
\mbox{minimize} &
\| RP_\pi^T\tilde\Theta - \tilde Y\|_F^2\\
\mbox{subject to} & \tilde \Theta~\mbox{upper triangular},
\end{array}
\EEAS
with variable $\tilde\Theta \in \reals^{p \times p}$. Here $\| \cdot \|_F^2$ is
the Frobenius norm squared, \ie, the sum of the entries.  The matrix $\tilde Y$
is given by $\tilde Y = \tilde y \ones^T$, where $\ones$ is the vector with all
entries one, \ie, $\tilde Y$ is the matrix with all columns $\tilde y$. (The
$p$ different least-squares problems are uncoupled, but it is convenient to
represent them as one matrix least-squares problem \cite{Boyd2018}.)

Let $\tilde Q \tilde R=RP_\pi^T$ denote the QR decomposition of $RP_\pi^T$.
Substituting $\tilde Q\tilde R$ for $RP_\pi^T$ above, and multiplying the
argument of the Frobenius norm the orthogonal matrix $\tilde Q^T$, the problem
above can be written as
\BEAS
\begin{array}{ll}
\mbox{minimize} &
\|\tilde R\tilde\Theta - \tilde Q^T \tilde Y\|_F^2\\
\mbox{subject to} & \tilde\Theta~\mbox{upper triangular},
\end{array}
\EEAS
with variable $\tilde\Theta \in \reals^{p \times p}$. The solution has the
simple form
\BEQ\label{e-Theta-star}
\tilde\Theta^\star = \tilde
R^{-1}\mathbf{triu}(\tilde Q^T \tilde Y).
\EEQ
where $\mathbf{triu}(\cdot )$ gives the upper triangular part of its argument,
\ie, sets the strictly lower triangular entries to zero. Note that the
righthand side is upper triangular since upper triangularity is preserved under
inversion and matrix multiplication. This result is equivalent to application
of the Frish--Waugh--Lovell theorem from econometrics
\cite{Frisch1933,Lovell1963} and is also well-known in statistics
\cite{Hastie2009}. The optimal parameters for $\mathcal S_0, \mathcal S_1,
\ldots, \mathcal S_p$ are thus the columns of
$\Theta^\star=P_\pi^T\tilde\Theta^\star$.

\paragraph{Complexity.}
Computing the QR factorization of $RP_\pi^T$ costs $O(p^3)$. We can form
$\tilde Q^T\tilde Y = \tilde Q^T\tilde y \ones$ in $O(p^2)$, which is
negligible. We can compute $\Theta^\star$ using \eqref{e-Theta-star} in
$O(p^3)$ flops. In other words: We can find the parameter vectors for a whole
chain in $O(p^3)$, the same cost as solving a single least-squares problem with
$p$ variables and $p$ equations. We evidently save a factor of $p$, compared to
the na\"{i}ve method of solving $p$ least-squares problems, which has cost
$O(p^4)$.

It is easily verified that the cost of evaluating the $p$ least-squares
parameters on the test data is also $O(p^3)$, so the cost of evaluating the
lifts for the chain is $O(p^3)$.

\subsection{Summary}
Altogether, the complexity of LS-SPA is \BEQ \label{e-complexity-lssa} O(Np^2 +
Kp^3), \EEQ which can be compared to the complexity of the na\"{i}ve method,
$O(KNp^3)$ \eqref{e-complexity-naive}. The speedup over the na\"{i}ve method is
at least the minimum of $N$ and $Kp$, neither of which is typically small. We
note that LS-SPA can also be parallelized, by computing the lifts for each
$\pi\in \Pi$ in parallel.

\subsection{Quasi-Monte Carlo approximation}\label{s-qmc} Here we explain an
improvement over the simple Monte Carlo method in \eqref{e-mc-approx}.  (This
improvement has nothing to do with the problems being least-squares and is
applicable in other cases.) We will use quasi-Monte Carlo (QMC) sampling
instead of randomly sampling permutations to obtain $\Pi$. One proposed method
(which we call \emph{permutohedron QMC}) is given in \cite{Mitchell2022}. It
maps a Sobol' sequence in $[0,1]^{p-2}$ onto the permutohedron for $p$-element
permutations by mapping to the $(p-1)$-sphere, then embedding the
$(p-1)$-sphere into $\reals^p$ via an area-preserving transform and rounding
points to the nearest permutohedron vertex.

We propose another method (which we call \emph{argsort QMC}), which is to take
a Sobol' sequence on $[0,1]^p \subset \reals^p$, and choose the permutations as
the argsort (permutation that gives the sorted ordering) of each point in the
sequence. We have found empirically that this method does as well as
permutohedron sampling for this problem, and is computationally simpler.

We note that other quasi-Monte Carlo sequences, such as Halton sequences,
may be used in place of Sobol' sequences. It is common to randomize quasi-Monte
Carlo sequences as doing so can improve convergence rates, and some theoretical
work exists to justify error estimation in this setting
\cite{Owen1998,Owen2023}.
In our empirical studies, we have found scrambled Sobol' sequences to work well.

\subsection{Risk estimation}\label{s-bootstrap}

A natural question is how large the number of sampled permutations $K$
needs to be to obtain an accurate estimate of the Shapley values. In this
section we provide a method to estimate the error in the Shapley attribution
approximations provided by LS-SPA.
The error estimates provide the user with an idea of the
precision to which the attributions are accurate and can be used in a stopping
criterion.

\paragraph{Error.} We define the error in the estimate of the $j$th Shapley
value to be
\BEQ\label{e-feature-error}
|\hat S_j - S_j|,
\EEQ
where $S \in \reals^p$ is the vector of true Shapley values and $\hat S
\in \reals^p$ is the vector of approximate Shapley values as described in
\S\ref{s-shapval}. We also define the overall error in the Shapley estimate to
be
\BEQ
\label{e-overall-error} \|\hat S-S\|_2.
\EEQ
Other error metrics can be used, but \eqref{e-overall-error} is a simple
and default metric.

\paragraph{Risk estimation.}
Since the exact value of $S$ is not known, \eqref{e-feature-error} and
\eqref{e-overall-error} cannot be computed exactly. But, we can
efficiently estimate the errors using the central limit theorem.

If a permutation $\pi$ is sampled from the uniform
distribution on $\mathcal{P}$, then the expected value of $L(\pi)$ is $S$. Let
$\Sigma$ denote the covariance of $L(\pi)$. The central limit theorem
guarantees that $\sqrt{K}(\hat S - S)$ converges in distribution to
$\mathcal{N}(0, \Sigma)$ as $K\to\infty$. We can thus estimate the $q$th
quantile values of \eqref{e-feature-error} and \eqref{e-overall-error} over the
distribution of $\hat S$ for $K$ samples via Monte Carlo. We take $\hat\Sigma$
to be the unbiased sample covariance of $\{L(\pi)\}_{\pi\in\Pi}$, and sample
$D$ vectors $\Delta^{(1)}, \ldots, \Delta^{(D)}$ from $\mathcal{N}(0,
\frac{1}{K}\hat\Sigma)$. We then report the estimated error for feature $j$ as
\BEAS
\hat\rho_j = \mathbf{quantile}(\{|\Delta^{(i)}_j|\}_{i=1}^D; q)
\EEAS
and the estimated overall error as
\BEAS
\hat\sigma = \mathbf{quantile}(\{\|\Delta^{(i)}\|_2\}_{i=1}^D; q),
\EEAS
where $\mathbf{quantile}(\cdot; q)$ denotes the $q$th quantile.
A higher value of $q$ provides a more conservative error estimate.

\paragraph{Batching.}
To use the risk estimate in a stopping criterion, we can compute $\hat S$ in
batches. After each batch, we recompute $\hat\sigma$ and terminate early if it
is below a fixed tolerance $\epsilon>0$. More precisely, we set a batch size
$B$, a maximum number of batches $K/B$, and a risk tolerance $\epsilon>0$.

For any subset $\Pi$ of permutations, define the sample mean
\BEQ\label{e-sample-mean}
\hat S(\Pi) = \frac{1}{|\Pi|}\sum_{\pi \in \Pi}L(\pi)
\EEQ
and the biased sample
covariance
\BEQ\label{e-sample-cov}
\hat\Sigma_b(\Pi) = \frac{1}{|\Pi|}\sum_{\pi\in\Pi}(L(\pi)-\hat
S(\Pi))(L(\pi)-\hat S(\Pi))^T.
\EEQ
Instead of computing $\Pi$, $\hat S$, and the risk estimate all at once, we
compute them iteratively via batches $\Pi^{(1)}, \ldots, \Pi^{(K/B)}$, each of
size $B$. Initialize the estimated Shapley values $\hat S^{(0)} = 0$ and the
estimated biased sample covariance $\hat\Sigma^{(0)}_b = 0$. In iteration $j$,
we can compute $\hat S^{(j)}$ using the update rule \BEQ\label{e-update} \hat
S^{(j)} = \frac{j-1}{j}\hat S^{(j-1)} + \frac{1}{j}\hat S(\Pi^{(j)}), \EEQ
which holds since $\Pi^{(1)}, \ldots, \Pi^{(K/B)}$ are equally sized. We can
also compute $\hat\Sigma^{(j)}_b$ using the update rule provided in
\cite{Schubert2018},
\BEQ\label{e-update-cov}
\begin{aligned}
	\hat \Sigma^{(j)}_b &=
	\frac{j-1}{j}\hat\Sigma^{(j-1)}_b + \frac{1}{j}\hat\Sigma_b(\Pi^{(j)})\\
	&\phantom{{}={}}+\frac{j-1}{j^2}D^{(j)}{D^{(j)}}^T,
\end{aligned}
\EEQ
where $D^{(j)}=\hat S^{(j-1)}-\hat S(\Pi^{(j)})$. The unbiased sample
covariance $\hat\Sigma^{(j)}$ is $\frac{jB}{jB-1}\hat\Sigma^{(j)}_b$, which we
can use to generate our risk estimates.

Note that batching in this manner can result in terminating early when $\hat S$
is computed on a number of permutations that is not a power of $2$. When using
Sobol' sequences, this can destroy the balance properties expected of QMC, but
in practice, we have found this does not matter.

The central limit theorem is based on random samples, which is not the case for
QMC methods. As a result, risk estimates when $\hat S$ is computed via a QMC
method to sample permutations do not come with the theoretical guarantees that
random samples have.  We have observed empirically that estimates using QMC are
still good estimates of the actual errors.

\subsection{Sample augmentation} Monte Carlo and QMC sampling
techniques can be augmented to potentially further reduce estimate variance. A
simple way to do this is via antithetical sampling, in which for each
permutation $\pi$ sampled, the permutation $\gamma\pi$ is also included, where
$\gamma$ is the permutation that reverses the sequence $1,\ldots,p$. The
permutation $\gamma\pi$ corresponds to the feature chain
\[
\emptyset \subset \{\pi_k\} \subset \{\pi_k,\pi_{k-1}\} \subset \cdots \subset \{\pi_k, \pi_{k-1}, \ldots, \pi_1\}.
\]
Note that if antithetical sampling is used, the sample mean
\eqref{e-sample-mean} should be adjusted as
\[
\hat S(\Pi) = \frac{1}{|\Pi|}\sum_{\pi\in\Pi}\tilde L(\pi),
\]
and the sample covariance \eqref{e-sample-cov} should be adjusted to
\[
\hat\Sigma_b(\Pi)=\frac{1}{|\Pi|}\sum_{\pi\in\Pi}(\tilde L(\pi)-\hat S(\Pi))
(\tilde L(\pi)-\hat S(\Pi)),
\]
where $\tilde L(\pi)=(L(\pi)+L(\gamma\pi))/2$. Empirically, we find that for
well-conditioned data, antithetical sampling works very well when combined with
Monte Carlo or QMC sampling and often converges more than twice as quickly.
However, for data with poorly conditioned empirical covariance matrices,
antithetical sampling gains little additional performance.

A more sophisticated technique is the ergodic sampling technique described in
\cite{Illes2022}, which increases the number of permutations by shuffling each
sampled permutation in a way that greedily minimizes the covariances of the
lift vectors. However, this technique applied here introduces an $O(p^5)$ cost,
so is not suitable for large $p$.

\subsection{Algorithm summary}\label{s-lssa}
The LS-SPA algorithm is summarized in algorithm \ref{a-lsspa}. We note that
in line~2, the Cholesky reduction described in \S\ref{s-large-data} may be used
instead of the QR reduction described in \S\ref{s-reduce}.
\begin{algorithm}\label{a-lsspa}
	\SetAlgoLined
	\LinesNumbered
	\KwResult{Return $\hat S$, $\{\hat\rho_i\}_{i=1}^p$, and $\hat \sigma$}
	\textbf{Given:} Training data $X\in \reals^{N\times p}$, training labels $y \in
	\reals^{N}$, test data $X^\text{tst}\in \reals^{M\times p}$, test labels $y^\text{tst} \in
	\reals^{M}$, maximum number of sampled permutations $K\in \integers_{++}$,
	batch size $B\in \integers_{++}$, risk tolerance $\epsilon\in \reals_{++}$,
	error quantile $q\in (0,1)$\;
	Reduce $X,y,X^\text{tst},y^\text{tst}$ as described in \S\ref{s-reduce}\;
	Generate $K$ permutations $\pi^{(1)},\ldots,\pi^{(K)}$ as described in
	\S\ref{s-shapval} or \S\ref{s-qmc}\;
	\For{$j=1,\ldots, K/B$}{
		\For{$k=(j-1)B+1,\ldots,jB$}{
			Compute lifts $L(\pi^{(k)})$ as described in \S\ref{s-reduce}, and if antithetical
			sampling is used, also compute $L(\gamma\pi^{(k)})$\;
		}
		Compute approximate attributions $\hat S^{(j)}$ and estimated overall errors
		$\hat\sigma$ as described in \S\ref{s-bootstrap}\;
		\If{$\textnormal{error estimate below tolerance, }\hat\sigma<\epsilon$}{
			Compute estimated feature errors $\{\hat\rho_i\}_{i=1}^p$ as described in
			\S\ref{s-bootstrap}\;
			\Return{$\hat S=\hat S^{(j)}$, $\{\hat\rho_i\}_{i=1}^p$, $\hat \sigma$}
		}
	}
	Print tolerance not reached warning\;
	Compute $\hat S^{(j)}$ and $\hat\sigma$ as described in \S\ref{s-bootstrap}\;
	Compute $\{\hat\rho_i\}_{i=1}^p$ as described in \S\ref{s-bootstrap}\;
	\Return{$\hat S=\hat S^{(K/B)}$, $\{\hat\rho_i\}_{i=1}^p$, $\hat \sigma$}\;
	\caption{{\sc Least-squares Shapley attribution (LS-SPA)} \label{a-lssa}}
\end{algorithm}


\subsection{Implementation}\label{s-implementation}
We have developed two Python implementations of algorithm \ref{a-lssa}. The
computational results we present in \S\ref{s-num} are derived from a JAX-based
\cite{Jax2018} implementation of algorithm \ref{a-lssa} and some of the
extensions discussed in \S\ref{s-ext}. The JAX implementation, along with our
numerical experiments, is available at
\begin{quote}
	\url{https://github.com/cvxgrp/ls-spa-benchmark}.
\end{quote}
We also provide a more user-friendly, NumPy-based \cite{Harris2020} library
implementing algorithm \ref{a-lssa} at
\begin{quote}
	\url{https://github.com/cvxgrp/ls-spa}.
\end{quote}

JAX allows LS-SPA to utilize GPU(s), while the NumPy implementation of
LS-SPA runs on CPUs only. In addition, the JAX implementation employs some
additional parallelization to execute LS-SPA more efficiently, although as a
consequence, the JAX implementation is harder to read and harder to use.
In addition, JAX is harder to install and configure, especially in order
to use GPU(s). For this reason, we provide the NumPy implementation, which lacks
some of the features (notably QMC) and performance of the JAX implementation,
but is in turn much easier to install, read, and use.

\section{Extensions and variations}\label{s-ext} In this section, we describe
some extensions to the basic problem and method described above.

\subsection{Cross validation metric}
In the discussion above we used simple out-of-sample validation, but we can
also use other more sophisticated validation methods, such as $M$-fold cross
validation \cite[Ch.~17]{Efron93}. Here the original data are split into $M$
different `folds'.  For $m=1, \ldots, M$ we fit a model using as training data
all folds except $m$ and validate it on fold $m$. We use the average validation
mean-square error to obtain the $R^2$ score. The methods above apply
immediately to this situation.

\subsection{Ridge regularization}
In ridge regression, we choose the parameter $\theta$ by solving the
$\ell_2$-regularized least-squares problem \BEQ\label{e-rlsq}
\begin{array}{ll}
	\mbox{minimize} & \frac{1}{N}\| X \theta - y\|_2^2 + \lambda\|\theta\|_2^2,
\end{array}
\EEQ where $\theta\in \reals^p$ is the optimization variable, $X \in \reals^{N
	\times p}$ and $y \in \reals^N$ are data, and $\lambda$ is a positive
regularization hyperparameter. Observe that \eqref{e-rlsq} can be reformulated
as \BEQ\label{e-blsq}
\begin{array}{ll}
	\mbox{minimize} & \| \tilde X \theta - \tilde y\|_2^2,
\end{array}
\EEQ where $\tilde X$ and $\tilde y$ are the stacked data
\BEAS
\tilde{X}=\begin{bmatrix}
	X/\sqrt{N} \\
	\sqrt{\lambda}I
\end{bmatrix}, \qquad
\tilde{y}=\begin{bmatrix}
	y/\sqrt{N} \\
	0
\end{bmatrix}.
\EEAS
This reformulation transforms the regularized problem \eqref{e-rlsq} into a
least-squares problem in the form of \eqref{e-lsq}. As such, we can now perform
LS-SPA on the regularized problem.

To choose the value of the hyper-parameter $\lambda$, we consider a set of
candidate values $\lambda_1, \ldots, \lambda_L$.  We solve the regularized
least-squares regression problem for each one and evaluate the resulting
parameter $\lambda$ using out-of-sample or cross-validation.  We then choose
$\lambda$ as the one among our choices that achieves the lower mean-square test
error.  We use this value to compute the $R^2$ metric.

\subsection{Very large data}\label{s-large-data} If $X$ is too large to fit
into memory such that performing the initial QR factorization cannot be done,
one alternative is to compute the Cholesky factorization of the covariance
matrix of $[X ~ y]$, \ie, the matrix
\BEAS
\hat\Sigma = \begin{bmatrix}X^T \\ y^T\end{bmatrix}
\begin{bmatrix}X & y\end{bmatrix}
= \begin{bmatrix}X^T X & X^T y \\ y^T X & y^T y\end{bmatrix}.
\EEAS
The covariance matrix $\hat\Sigma$ is $p \times p$ and can be computed via
block matrix multiplication by blocking $[X \, y]$ vertically, making it
possible to distribute the computation across multiple devices or compute
iteratively on one device. The upper-triangular factor $\tilde R$ in the
Cholesky factorization $\tilde R^T \tilde R=\hat\Sigma$ can then be blocked as
\BEAS
\tilde R = \begin{bmatrix}
	R & Q^T y \\
	0 & \|y - Q(Q^T y)\|_2
\end{bmatrix}
\EEAS
where $QR=X$ is the QR factorization of $X$. We can thus extract $R$, $Q^T y$,
and $\|y - Q(Q^T y)\|_2$ from $\tilde R$ to compute the reduction
\eqref{e-reduction} for use in LS-SPA. This alternative approach costs
$O(Np^2)$ flops for the computation of $\hat\Sigma$ and $O(p^3)$ flops for the
computation of $\tilde R$, giving a total cost of $O(Np^2)$, the same as the QR
method. However, Cholesky factorization is less stable than QR and can fail
for poorly conditioned $\hat\Sigma$.

\subsection{Non-quadratic regularizers}
We consider the case where the quadratic loss is paired with a non-quadratic
but convex regularizer.  This means we choose the model parameter $\theta$ by
solving \BEQ\label{e-nonquad}
\begin{array}{ll}
	\mbox{minimize} & \|X\theta -y\|_2^2 + \lambda r(\theta) \end{array}, \EEQ with
variable $\theta\in \reals^p$, data $X\in \reals^{N\times p}$ and $y\in
\reals^N$, and convex but non-quadratic regularizer $r: \reals^p \to \reals
\cup \{\infty\}$.  Here $\lambda$ is the regularization hyper-parameter. Simple
examples include the nonnegative indicator function, so the problem above is a
non-negative least-squares problem.  Another example is
$r(\theta)=\|\theta\|_1$, which gives the lasso problem \cite{Hastie2009}.

While our formula for $\theta$ given in \S\ref{s-reduce} no longer holds, we
can still reduce the complexity of the computation with the initial reduction.
Thus when we find $\theta$ we solve a smaller convex optimization problem with
a square data matrix.

\section{Numerical experiments}\label{s-num}

\subsection{Experiment descriptions}\label{s-num-desc}
We describe two numerical experiments, one medium size and one large, that
demonstrate the relationship between the runtime of the LS-SPA and the accuracy
of the approximated Shapley attribution. The code for the experiments can be
found in
\begin{quote}
	\url{https://github.com/cvxgrp/ls-spa-benchmark}.
\end{quote}

\paragraph{Medium size experiment.} The medium-size experiment uses a single
randomly generated data set with $p=100$ features and $N=M=10^5$ data points
for the train and test data sets. Details of data generation are given in
\S\ref{s-data-gen}. The medium-size experiment is meant to show how the overall
error in the estimate of the Shapley attributions evolves with an increasing
number of sampled feature chains.
We run LS-SPA once with of each of three
methods to sample feature chains (MC, permutohedron QMC, and argsort QMC) in
the medium size experiment, with and without antithetical sampling, for a total
of $K=2^{13}$ sampled permutations. For the QMC methods, we use scrambled
Sobol' sequences with SciPy's default scrambling strategy, which is a (left)
linear matrix scramble followed by a digital random shift
\cite{Virtanen2020,Matousek1998}. We track the estimated overall error and the
true overall error as more permutations are sampled during the runtime of
LS-SPA. We compare the true overall errors achieved by each sampling method,
and we compare the error estimate to the true error for MC and argsort QMC. For
the purpose of computing the true overall error, we compute the
``ground-truth'' Shapley attributions by running LS-SPA with Monte Carlo and
antithetical sampling for $2^{28}$ total permutations. The quantile we use for
risk estimation is $q = 0.95$.

\paragraph{Large experiment.} The large experiment uses a single randomly
generated data set with $p=1000$ features and $N=M=10^6$ data points for the
train and test data sets. Details of data generation are given in
\S\ref{s-data-gen}. The large experiment is a timing test meant to demonstrate
that LS-SPA scales to large problems. The large experiment uses a single
run of LS-SPA with Monte Carlo and antithetical sampling only and is run until
the error estimate falls below a tolerance $\epsilon=10^{-3}$. The quantile we
use for risk estimation is $q = 0.95$.

\paragraph{Computation platforms.} The medium-size experiment, except for
computation of the ground-truth Shapley attributions, was done on a 16-thread
Intel Core i7-10875H CPU at 2.30~GHz with 64~GB RAM. The large experiment and
computation of ground truth for the medium experiment were done with two
Intel Xeon E5-2640 v4 CPUs, each with 20 threads, and four
NVIDIA GTX TITAN X GPUs, each of which has 12~GB RAM. Note that for the
large experiment and computation of ground truth for the medium experiment,
all numerical computations were done on GPU.

\subsection{Data generation}\label{s-data-gen} For both experiments, we solved
instances of \eqref{e-lsq} on randomly generated train and test data,
$(X^\mathrm{trn},y^\mathrm{trn})$ and $(X^\mathrm{tst},y^\mathrm{tst})$,
respectively. To generate the data, we first randomly generate a feature
covariance matrix $\Sigma = FF^T + I$, where $F \in \reals^{p \times (p/20)}$
is generated by sampling its entries independently from a $\mathcal N(0, 1)$
distribution. We then let $C$ be the correlation matrix of $\Sigma$.

Next, the true vector of feature coefficients $\theta$ was generated by
randomly selecting $\lfloor (p+1)/10\rfloor$ entries to be $2$ and the
remaining entries to be $0$.

Finally, we generate $X^\mathrm{trn} \in \reals^{N\times p}$ and
$X^\mathrm{tst}\in\reals^{M\times p}$, consisting, respectively, of $N$ and $M$
observations generated independently at random from a $\mathcal N(0, C)$
distribution. We then generate noise vectors $\omega^\mathrm{trn},
\omega^\mathrm{tst} \in \reals^p$ independently from a $\mathcal N(0, (3p^2/2)
I)$ distribution and construct $y^\mathrm{trn} = X^\mathrm{trn}\theta +
\omega^\mathrm{trn}$ and $y^\mathrm{tst} = X^\mathrm{tst}\theta +
\omega^\mathrm{tst}$. Finally, we include an intercept in our linear model by
centering the columns of $X^\mathrm{trn}$ and $X^\mathrm{tst}$ by subtracting
the respective column means of $X^\mathrm{trn}$, and also centering
$y^\mathrm{trn}$ and $y^\mathrm{tst}$ by subtracting the mean of
$y^\mathrm{trn}$, as discussed in \S\ref{s-least-squares}.
The features generated had high correlation, which we found empirically was
adversarial for LS-SPA.

\subsection{Results}
\begin{figure}[t]
	\centering
	\includegraphics[width=0.65\linewidth]{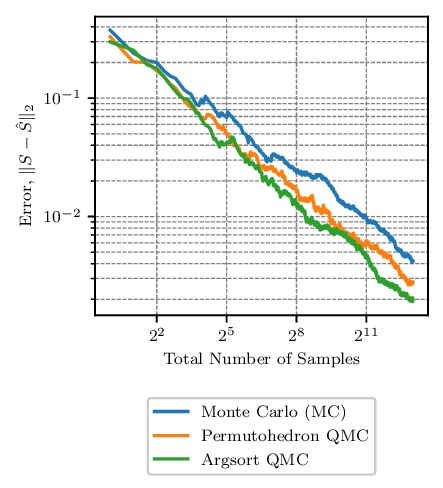}
	\caption{Overall error versus number of samples on the medium-size dataset
		using MC (blue), permutohedron QMC (orange), and argsort QMC (green) without
		antithetical sampling to sample feature chains.}
	\label{f-err}
\end{figure}

\begin{figure}[t]
	\centering
	\includegraphics[width=0.65\linewidth]{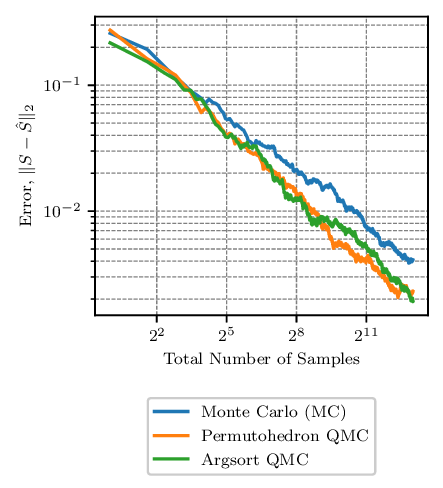}
	\caption{Overall error versus number of samples on the medium-size dataset
		using MC (blue), permutohedron QMC (orange), and argsort QMC (green) with
		antithetical sampling to sample feature chains.}
	\label{f-err-antithetical}
\end{figure}

\paragraph{Medium size experiment.} We used each of MC, permutohedron QMC,
and argsort QMC, with and without antithetical sampling, to sample $K = 2^{13}$
total feature chains, done in $2^5$ batches of size $2^8$ to illustrate the
progress of LS-SPA as more permutations are sampled.  LS-SPA took an average
of 9 minutes 12 seconds to compute $2^{13}$ lift vectors, which is 67.4
milliseconds per lift vector. In comparison, a na\"ive implementation that does
not take advantage of any reductions described in LS-SPA took 4 minutes 26
seconds to compute $2^3$ lift vectors, which is 33.3 seconds per lift vector.
The errors for each method sampling method as a function of the number of
feature chains completed are shown in figures \ref{f-err} and
\ref{f-err-antithetical}. Note that the condition number of $C$ was $316.0$.

\begin{figure}[t]
	\centering
	\includegraphics[width=0.65\linewidth]{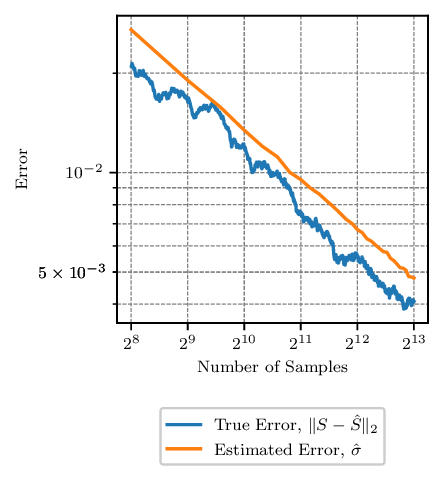}
	\caption{True error (blue) and estimated error (orange) while running LS-SPA
		using Monte Carlo with antithetical sampling to sample feature chains.}
	\label{f-risk-mc}
\end{figure}

\begin{figure}[t]
	\centering
	\includegraphics[width=0.65\linewidth]{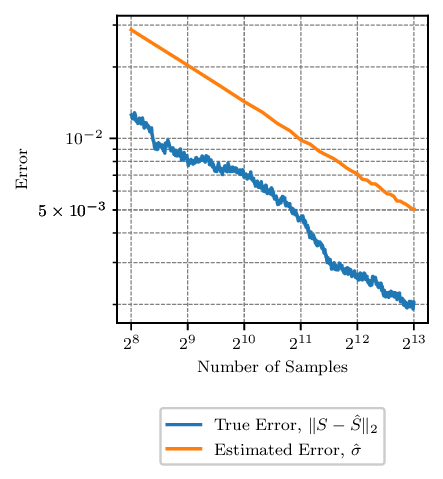}
	\caption{True error (blue) and estimated error (orange) while running LS-SPA
		using argsort QMC without antithetical sampling to sample feature chains.}
	\label{f-risk-argsort}
\end{figure}

In figure \ref{f-risk-mc}, we also plot the true overall error against the
error estimate, which was computed using the risk estimation procedure
described in \S\ref{s-bootstrap}, at each step of the algorithm using Monte
Carlo with antithetical sampling to sample feature chains. We plot the same
things in figure \ref{f-risk-argsort} using argsort QMC without antithetical
sampling to sample feature chains.

\paragraph{Large experiment} We use Monte Carlo with antithetical sampling
to sample feature chains and run LS-SPA until the estimated error $\hat\sigma$
is below the tolerance level $\epsilon=10^{-3}$. We use a quantile value of
$q=0.95$. Since the data were too large to fit on one GPU, we use the Cholesky
reduction presented in \S\ref{s-large-data}.

The algorithm took 27.2 seconds to complete the initial reduction. LS-SPA ran
for 230.4 seconds to reach an error estimate of $9.9\times 10^{-4}$, computing
a total of 29,696 lift vectors, done in 29 batches of $2^8$ permutations on
each of the four GPUs. This gives an average of $7.8\times 10^{-3}$ seconds per
lift vector. Note that the correlation matrix $C$ used to generate the data has
condition number $2.4 \times 10^3$.

\subsection{Discussion} For moderately sized $p$, \emph{e.g.}, on the order of
$100$, the NumPy implementation of LS-SPA linked in \S\ref{s-implementation}
fairly quickly converges to an estimate of the Shapley attributions with error
$10^{-3}$. This is true even when using Monte Carlo sampling, which tends to
underperform compared to quasi-Monte Carlo sampling techniques. For $p$ of this
size, LS-SPA achieves a $500\times$ speedup compared to a na\"ive estimation
procedure.

For larger $p$ on the order of $1000$ or more, the NumPy implementation can
still be used, but a more specialized implementation should be used for maximum
performance. An example JAX implementation is available in the benchmark repo
linked in \S\ref{s-num-desc}.

\subsection*{Acknowledgments}

We thank Ron Kahn for suggesting the topic, Kunal Menda for recommending the
use of quasi-Monte Carlo, Trevor Hastie and Emmanuel Cand\`es for suggesting
the risk estimation method, and Art Owen and Thomas Schmelzer for helpful
feedback on a draft.  We are indebted to kind readers for pointing us to
literature we had missed and in addition suggesting methods such as
antithetical sampling.

\clearpage
{\small
	\bibliography{refs.bib}

\newcommand{\etalchar}[1]{$^{#1}$}
\begin{thebibliography}{HMvdW{\etalchar{+}}20}

\bibitem[AFSS19]{Algaba2019}
Encarnaci{\'{o}}n Algaba, Vito Fragnelli, and Joaqu{\'{i}}n S{\'{a}}nchez-Soriano.
\newblock {\em Handbook of the {S}hapley Value}.
\newblock Chapman \& Hall/CRC, Boca Raton, Florida, USA, 2019.

\bibitem[BFH{\etalchar{+}}23]{Jax2018}
James Bradbury, Roy Frostig, Peter Hawkins, Matthew~James Johnson, Chris Leary, Dougal Maclaurin, George Necula, Adam Paszke, Jake Vander{P}las, Skye Wanderman-{M}ilne, and Qiao Zhang.
\newblock {JAX}: {C}omposable transformations of {P}ython+{N}um{P}y programs, 2023.

\bibitem[BG65]{Businger1965}
Peter Businger and Gene Golub.
\newblock Linear least squares solutions by householder transformations.
\newblock {\em Numerische Mathematik}, 7(3):269--276, June 1965.

\bibitem[Bud93]{Budescu1993}
David Budescu.
\newblock Dominance analysis: {A} new approach to the problem of relative importance of predictors in multiple regression.
\newblock {\em Psychological Bulletin}, 114(3):542--551, 1993.

\bibitem[BV18]{Boyd2018}
Stephen Boyd and Lieven Vandenberghe.
\newblock {\em Introduction to Applied Linear Algebra: {V}ectors, Matrices, and Least Squares}.
\newblock Cambridge University Press, Cambridge, UK, 2018.

\bibitem[CCLL23]{Chen2023}
Hugh Chen, Ian Covert, Scott Lundberg, and Su-In Lee.
\newblock Algorithms to estimate {S}hapley value feature attributions.
\newblock {\em Nature Machine Intelligence}, 5(6):590--601, May 2023.

\bibitem[CEW12]{Chalkiadakis2012}
Georgios Chalkiadakis, Edith Elkind, and Michael Wooldridge.
\newblock {\em Computational Aspects of Cooperative Game Theory}.
\newblock Springer International Publishing, Cham, 2012.

\bibitem[CGMT17]{Castro2017}
Javier Castro, Daniel G{\'{o}}mez, Elisenda Molina, and Juan Tejada.
\newblock Improving polynomial estimation of the {S}hapley value by stratified random sampling with optimum allocation.
\newblock {\em Computers {\&} Operations Research}, 82:180--188, June 2017.

\bibitem[CGT09]{Castro2009}
Javier Castro, Daniel G{\'{o}}mez, and Juan Tejada.
\newblock Polynomial calculation of the {S}hapley value based on sampling.
\newblock {\em Computers {\&} Operations Research}, 36(5):1726--1730, May 2009.

\bibitem[CPT89]{Curiel1989}
Imma Curiel, Giorgio Pederzoli, and Stef Tijs.
\newblock Sequencing games.
\newblock {\em European Journal of Operational Research}, 40(3):344--351, June 1989.

\bibitem[CS91]{Chevan1991}
Albert Chevan and Michael Sutherland.
\newblock Hierarchical partitioning.
\newblock {\em The American Statistician}, 45(2):90--96, May 1991.

\bibitem[DNW81]{Dubey1981}
Pradeep Dubey, Abraham Neyman, and Robert Weber.
\newblock Value theory without efficiency.
\newblock {\em Mathematics of Operations Research}, 6(1):122--128, 1981.

\bibitem[DP94]{Deng1994}
Xiaotie Deng and Christos Papadimitriou.
\newblock On the complexity of cooperative solution concepts.
\newblock {\em Mathematics of Operations Research}, 19(2):257--266, May 1994.

\bibitem[ET93]{Efron93}
Bradley Efron and Robert Tibshirani.
\newblock {\em An Introduction to the Bootstrap}.
\newblock Number~57 in Monographs on Statistics and Applied Probability. Chapman \& Hall/CRC, Boca Raton, Florida, USA, 1993.

\bibitem[FAB{\etalchar{+}}02]{Fernandez2002}
Julio Fern{\'{a}}ndez, Encarnaci\'on Algaba, Jes\'us Bilbao, Andr\'es Jim{\'{e}}nez, Nerecsy Jim{\'{e}}nez, and Jorge L{\'{o}}pez.
\newblock Generating functions for computing the {M}yerson value.
\newblock {\em Annals of Operations Research}, 109(1/4):143--158, 2002.

\bibitem[FK92]{Faigle1992}
Ulrich Faigle and Walter Kern.
\newblock The {S}hapley value for cooperative games under precedence constraints.
\newblock {\em International Journal of Game Theory}, 21(3):249--266, September 1992.

\bibitem[FSN21]{Fryer2021}
Daniel Fryer, Inga Strümke, and Hien Nguyen.
\newblock {S}hapley values for feature selection: {T}he good, the bad, and the axioms.
\newblock {\em IEEE Access}, 9:144352--144360, 2021.

\bibitem[FW33]{Frisch1933}
Ragnar Frisch and Frederick Waugh.
\newblock Partial time regressions as compared with individual trends.
\newblock {\em Econometrica}, 1:387, 1933.

\bibitem[FWJ08]{Fatima2008}
Shaheen Fatima, Michael Wooldridge, and Nicholas Jennings.
\newblock A linear approximation method for the {S}hapley value.
\newblock {\em Artificial Intelligence}, 172(14):1673--1699, 2008.

\bibitem[GKC02]{Granot2002}
Daniel Granot, Jeroen Kuipers, and Sunil Chopra.
\newblock Cost allocation for a tree network with heterogeneous customers.
\newblock {\em Mathematics of Operations Research}, 27(4):647--661, November 2002.

\bibitem[Gol65]{Golub1965}
Gene Golub.
\newblock Numerical methods for solving linear least squares problems.
\newblock {\em Numerische Mathematik}, 7(3):206--216, June 1965.

\bibitem[Gr{\"{o}}06]{Grmping2006}
Ulrike Gr{\"{o}}mping.
\newblock Relative importance for linear regression in {R}: {T}he package relaimpo.
\newblock {\em Journal of Statistical Software}, 17(1), 2006.

\bibitem[Gr{\"{o}}15]{Grmping2015}
Ulrike Gr{\"{o}}mping.
\newblock Variable importance in regression models.
\newblock {\em {WIREs} Computational Statistics}, 7(2):137--152, February 2015.

\bibitem[GZ19]{Ghorbani2019}
Amirata Ghorbani and James Zou.
\newblock Data {S}hapley: {E}quitable valuation of data for machine learning.
\newblock In {\em Proceedings of the 36th International Conference on Machine Learning}, volume~97 of {\em Proceedings of Machine Learning Research}, pages 2242--2251, Long Beach, CA, USA, 09--15 Jun 2019. PMLR.

\bibitem[HMvdW{\etalchar{+}}20]{Harris2020}
Charles~R. Harris, K.~Jarrod Millman, St{\'{e}}fan~J. van~der Walt, Ralf Gommers, Pauli Virtanen, David Cournapeau, Eric Wieser, Julian Taylor, Sebastian Berg, Nathaniel~J. Smith, Robert Kern, Matti Picus, Stephan Hoyer, Marten~H. van Kerkwijk, Matthew Brett, Allan Haldane, Jaime~Fern{\'{a}}ndez del R{\'{i}}o, Mark Wiebe, Pearu Peterson, Pierre G{\'{e}}rard-Marchant, Kevin Sheppard, Tyler Reddy, Warren Weckesser, Hameer Abbasi, Christoph Gohlke, and Travis~E. Oliphant.
\newblock Array programming with {NumPy}.
\newblock {\em Nature}, 585(7825):357--362, September 2020.

\bibitem[HPR22]{Harris2022}
Chris Harris, Richard Pymar, and Colin Rowat.
\newblock Joint {S}hapley values: a measure of joint feature importance.
\newblock In {\em International Conference on Learning Representations}, 2022.

\bibitem[HS12]{Huettner2012}
Frank Huettner and Marco Sunder.
\newblock Axiomatic arguments for decomposing goodness of fit according to {S}hapley and {O}wen values.
\newblock {\em Electronic Journal of Statistics}, 6:1239--1250, 2012.

\bibitem[HTF09]{Hastie2009}
Trevor Hastie, Robert Tibshirani, and Jerome Friedman.
\newblock {\em The Elements of Statistical Learning}.
\newblock Springer New York, New York City, NY, USA, 2009.

\bibitem[IK22]{Illes2022}
Ferenc Ill\'{e}s and P\'{e}ter Ker\'{e}nyi.
\newblock Estimation of the {S}hapley value by ergodic sampling, 2022.

\bibitem[IS05]{Ieong2005}
Samuel Ieong and Yoav Shoham.
\newblock Marginal contribution nets.
\newblock In {\em Proceedings of the 6th {ACM} conference on Electronic commerce}, Vancouver, British Columbia, Canada, June 2005. {ACM}.

\bibitem[K\'07]{Koczy2007}
L\'{a}szl\'o K\'{o}czy.
\newblock A recursive core for partition function form games.
\newblock {\em Theory and Decision}, 63(1):41--51, 2007.

\bibitem[Kru87]{Kruskal1987}
William Kruskal.
\newblock Relative importance by averaging over orderings.
\newblock {\em The American Statistician}, 41:6--10, 1987.

\bibitem[KVSF20]{Kumar2020}
I.~Elizabeth Kumar, Suresh Venkatasubramanian, Carlos Scheidegger, and Sorelle Friedler.
\newblock Problems with {S}hapley-value-based explanations as feature importance measures.
\newblock In {\em Proceedings of the 37th International Conference on Machine Learning}, volume 119 of {\em Proceedings of Machine Learning Research}, pages 5491--5500, online, 13--18 Jul 2020. PMLR.

\bibitem[LC01]{Lipovetsky2001}
Stan Lipovetsky and Michael Conklin.
\newblock Analysis of regression in game theory approach.
\newblock {\em Applied Stochastic Models in Business and Industry}, 17(4):319--330, 2001.

\bibitem[Lee03]{Leech2003}
Dennis Leech.
\newblock Computing power indices for large voting games.
\newblock {\em Management Science}, 49(6):831--837, June 2003.

\bibitem[LL17]{Lundberg2017}
Scott Lundberg and Su-In Lee.
\newblock A unified approach to interpreting model predictions.
\newblock In {\em Advances in Neural Information Processing Systems 30}, pages 4765--4774. Curran Associates, Inc., Long Beach, CA, USA, 2017.

\bibitem[LMG80]{Lindeman1980}
Richard Lindeman, Peter Merenda, and Ruth Gold.
\newblock {\em Introduction to Bivariate and Multivariate Analysis}.
\newblock Scott Foresman, Northbrook, IL, USA, 1980.

\bibitem[LO73]{Littlechild1973}
Stephen Littlechild and Guilliermo Owen.
\newblock A simple expression for the {S}hapley value in a special case.
\newblock {\em Management Science}, 20(3):370--372, 1973.

\bibitem[Lov63]{Lovell1963}
Michael Lovell.
\newblock Seasonal adjustment of economic time series and multiple regression analysis.
\newblock {\em Journal of the American Statistical Association}, 58(304):993--1010, December 1963.

\bibitem[MAS{\etalchar{+}}13]{Michalak2013E}
Tomasz Michalak, Karthik Aadithya, Piotr Szczepanski, Balaraman Ravindran, and Nicholas Jennings.
\newblock Efficient computation of the {S}hapley value for game-theoretic network centrality.
\newblock {\em Journal of Artificial Intelligence Research}, 46:607--650, April 2013.

\bibitem[Mat98]{Matousek1998}
Ji{\v r\'i} Matou{\v s}ek.
\newblock On the {$L_2$}-discrepancy for anchored boxes.
\newblock {\em Journal of Complexity}, 14:527--556, 12 1998.

\bibitem[MBA22]{Moehle2022}
Nicholas Moehle, Stephen Boyd, and Andrew Ang.
\newblock Portfolio performance attribution via {S}hapley value.
\newblock {\em Journal of Investment Management}, 20(3):33--52, 2022.

\bibitem[MCFH22]{Mitchell2022}
Rory Mitchell, Joshua Cooper, Eibe Frank, and Geoffrey Holmes.
\newblock Sampling permutations for {S}hapley value estimation.
\newblock {\em Journal of Machine Learning Research}, 23(43):1--46, 2022.

\bibitem[Mis16]{Mishra2016}
Sudhanshu Mishra.
\newblock {S}hapley value regression and the resolution of multicollinearity.
\newblock {\em {SSRN} Electronic Journal}, 2016.

\bibitem[Mol22]{Molnar2022}
Christoph Molnar.
\newblock Interpretable machine learning, 2022.

\bibitem[MP08]{Moretti2008}
Stefano Moretti and Fioravante Patrone.
\newblock Transversality of the {S}hapley value.
\newblock {\em {TOP}}, 16(1):1--41, April 2008.

\bibitem[MRS{\etalchar{+}}13]{Michalak2013}
Tomasz Michalak, Talal Rahwan, Piotr Szczepanski, Oskar Skibski, Ramasuri Narayanam, Nicholas Jennings, and Michael Wooldridge.
\newblock Computational analysis of connectivity games with applications to the investigation of terrorist networks.
\newblock In {\em International Joint Conference on Artificial Intelligence}, 2013.

\bibitem[MS60]{Mann1960}
Irwin Mann and Lloyd Shapley.
\newblock {\em Values of Large Games, IV: Evaluating the Electoral College by Montecarlo Techniques}.
\newblock RAND Corporation, Santa Monica, CA, 1960.

\bibitem[MS02]{Monderer2002}
Dov Monderer and Dov Samet.
\newblock Variations on the {S}hapley value.
\newblock In {\em Handbook of Game Theory with Economic Applications Volume 3}, volume~3 of {\em Handbook of Game Theory with Economic Applications}, chapter~54, pages 2055--2076. Elsevier, Amsterdam, Netherlands, 2002.

\bibitem[MTTH{\etalchar{+}}14]{Maleki2014}
Sasan Maleki, Long Tran-Thanh, Greg Hines, Talal Rahwan, and Alex Rogers.
\newblock Bounding the estimation error of sampling-based {S}hapley value approximation.
\newblock \url{https://arxiv.org/abs/1306.4265}, 2014.

\bibitem[OP17]{Owen2017}
Art Owen and Cl\'{e}mentine Prieur.
\newblock On {S}hapley value for measuring importance of dependent inputs.
\newblock {\em SIAM/ASA Journal on Uncertainty Quantification}, 5(1):986--1002, 2017.

\bibitem[Owe72]{Owen1972}
Guillermo Owen.
\newblock Multilinear extensions of games.
\newblock {\em Management Science}, 18(5):64--79, 1972.

\bibitem[Owe77]{Owen1977}
Guillermo Owen.
\newblock Values of games with a priori unions.
\newblock In {\em Mathematical Economics and Game Theory}, pages 76--88, Berlin, Heidelberg, 1977. Springer Berlin Heidelberg.

\bibitem[Owe98]{Owen1998}
Art~B. Owen.
\newblock Scrambling sobol' and niederreiter--xing points.
\newblock {\em Journal of Complexity}, 14:466--489, 12 1998.

\bibitem[Owe23]{Owen2023}
Art~B. Owen.
\newblock {\em Practical Quasi-Monte Carlo Integration}.
\newblock \url{https://artowen.su.domains/mc/practicalqmc.pdf}, 2023.

\bibitem[Pow07]{Powers2007}
Michael Powers.
\newblock Using {A}umann--{S}hapley values to allocate insurance risk.
\newblock {\em North American Actuarial Journal}, 11(3):113--127, 2007.

\bibitem[SG18]{Schubert2018}
Erich Schubert and Michael Gertz.
\newblock Numerically stable parallel computation of (co-)variance.
\newblock In {\em Proceedings of the 30th International Conference on Scientific and Statistical Database Management}, SSDBM '18, New York, NY, USA, 2018. Association for Computing Machinery.

\bibitem[Sha52]{Shapley1952}
Lloyd Shapley.
\newblock A value for {$N$}-person games.
\newblock In {\em Contributions to the Theory of Games ({AM}-28), Volume {II}}, pages 307--318. Princeton University Press, Princeton, NJ, USA, December 1952.

\bibitem[Stu92]{Stufken1992}
John Stufken.
\newblock Letters to the editor: On hierarchical partitioning.
\newblock {\em The American Statistician}, 46(1):70--77, 1992.

\bibitem[vCHHL17]{vanCampen2017}
Tjeerd van Campen, Herbert Hamers, Bart Husslage, and Roy Lindelauf.
\newblock A new approximation method for the {S}hapley value applied to the {WTC} 9/11 terrorist attack.
\newblock {\em Social Network Analysis and Mining}, 8(1), December 2017.

\bibitem[VGO{\etalchar{+}}20]{Virtanen2020}
Pauli Virtanen, Ralf Gommers, Travis~E. Oliphant, Matt Haberland, Tyler Reddy, David Cournapeau, Evgeni Burovski, Pearu Peterson, Warren Weckesser, Jonathan Bright, Stéfan~J. van~der Walt, Matthew Brett, Joshua Wilson, K.~Jarrod Millman, Nikolay Mayorov, Andrew R.~J. Nelson, Eric Jones, Robert Kern, Eric Larson, C~J Carey, İlhan Polat, Yu~Feng, Eric~W. Moore, Jake VanderPlas, Denis Laxalde, Josef Perktold, Robert Cimrman, Ian Henriksen, E.~A. Quintero, Charles~R. Harris, Anne~M. Archibald, Antônio~H. Ribeiro, Fabian Pedregosa, Paul van Mulbregt, Aditya Vijaykumar, Alessandro~Pietro Bardelli, Alex Rothberg, Andreas Hilboll, Andreas Kloeckner, Anthony Scopatz, Antony Lee, Ariel Rokem, C.~Nathan Woods, Chad Fulton, Charles Masson, Christian Häggström, Clark Fitzgerald, David~A. Nicholson, David~R. Hagen, Dmitrii~V. Pasechnik, Emanuele Olivetti, Eric Martin, Eric Wieser, Fabrice Silva, Felix Lenders, Florian Wilhelm, G.~Young, Gavin~A. Price, Gert-Ludwig Ingold, Gregory~E. Allen, Gregory~R. Lee, Hervé Audren, Irvin Probst, Jörg~P. Dietrich, Jacob Silterra, James~T Webber, Janko Slavič, Joel Nothman, Johannes Buchner, Johannes Kulick, Johannes~L. Schönberger, José~Vinícius de~Miranda~Cardoso, Joscha Reimer, Joseph Harrington, Juan Luis~Cano Rodríguez, Juan Nunez-Iglesias, Justin Kuczynski, Kevin Tritz, Martin Thoma, Matthew Newville, Matthias Kümmerer, Maximilian Bolingbroke, Michael Tartre, Mikhail Pak, Nathaniel~J. Smith, Nikolai Nowaczyk, Nikolay Shebanov, Oleksandr Pavlyk, Per~A. Brodtkorb, Perry Lee, Robert~T. McGibbon, Roman Feldbauer, Sam Lewis, Sam Tygier, Scott Sievert, Sebastiano Vigna, Stefan Peterson, Surhud More, Tadeusz Pudlik, Takuya Oshima, Thomas~J. Pingel, Thomas~P. Robitaille, Thomas Spura, Thouis~R. Jones, Tim Cera, Tim Leslie, Tiziano Zito, Tom Krauss, Utkarsh Upadhyay, Yaroslav~O. Halchenko, and Yoshiki Vázquez-Baeza.
\newblock Scipy 1.0: fundamental algorithms for scientific computing in python.
\newblock {\em Nature Methods}, 17:261--272, 3 2020.

\bibitem[WF20]{Williamson2020}
Brian Williamson and Jean Feng.
\newblock Efficient nonparametric statistical inference on population feature importance using shapley values.
\newblock In {\em Proceedings of the 37th International Conference on Machine Learning}, volume 119 of {\em Proceedings of Machine Learning Research}, pages 10282--10291, online, 13--18 Jul 2020. PMLR.

\bibitem[WZKG20]{Wang2020}
Jianhong Wang, Yuan Zhang, Tae-Kyun Kim, and Yunjie Gu.
\newblock {S}hapley {$Q$}-value: {A} local reward approach to solve global reward games.
\newblock {\em Proceedings of the {AAAI} Conference on Artificial Intelligence}, 34(05):7285--7292, April 2020.

\bibitem[ZMB18]{Zhao2018}
Kaifeng Zhao, Seyed~Hanif Mahboobi, and Saeed Bagheri.
\newblock {S}hapley value methods for attribution modeling in online advertising.
\newblock \url{https://arxiv.org/abs/1804.05327}, 2018.

\bibitem[ZR94]{Zlotkin1994}
Gilad Zlotkin and Jeffrey Rosenschein.
\newblock Coalition, cryptography, and stability: {M}echanisms for coalition formation in task oriented domains.
\newblock In {\em Proceedings of the Twelfth AAAI National Conference on Artificial Intelligence}, AAAI'94, pages 432--437, Seattle, WA, USA, 1994. AAAI Press.

\bibitem[ZSGJ23]{Zhang2023}
Haoran Zhang, Harvineet Singh, Marzyeh Ghassemi, and Shalmali Joshi.
\newblock "{W}hy did the model fail?": {A}ttributing model performance changes to distribution shifts.
\newblock In {\em Proceedings of the 40th International Conference on Machine Learning}, volume 202 of {\em Proceedings of Machine Learning Research}, pages 41550--41578, Honolulu, HI, USA, 23--29 Jul 2023. PMLR.

\end{thebibliography}
}

\end{document}